\documentclass[aps,amsfonts,epsf]{revtex4}
\usepackage{graphics}
\usepackage{graphicx}
\usepackage{epsfig}

\def\beq{\begin{equation}}
\def\eeq{\end{equation}}
\def\bea{\begin{eqnarray}}
\def\eea{\end{eqnarray}}

\def\reff@jnl#1{{\rm#1\/}}
\def\prd{\reff@jnl{Phys. Rev. D }}        
\def\cqg{\reff@jnl{Class. Quantum Grav. }} 
\def\ekp{Porter~E~K }
\def\njc{Cornish~N~J }
\def\jrg{Gair~J~R }
\begin{document}
\input epsf.tex

\title{A New Method of Accelerated Bayesian Inference for Comparable Mass Binaries in both Ground and Space-Based Gravitational Wave Astronomy.}
\author{Edward K. Porter} 
\affiliation{Fran\c{c}ois Arago Centre, APC, Universit\'e Paris Diderot,\\ CNRS/IN2P3, CEA/Irfu, Observatoire de Paris, Sorbonne Paris Cit\'e, \\10 rue A. Domon et L. Duquet, 75205 Paris Cedex 13, France\footnote{email:porter@apc.univ-paris7.fr}}
\vspace{1cm}

\vspace{1cm}
\begin{abstract}
With the advance in computational resources, Bayesian inference is increasingly becoming the standard tool of practise in GW astronomy.  However, algorithms
such as Markov Chain Monte Carlo (MCMC) require a large number of iterations to guarantee convergence to the target density.  Each chain demands a large
number of evaluations of the likelihood function, and in the case of a Hessian MCMC, calculations of the Fisher information matrix for use as a proposal 
distribution.  As each iteration requires the generation of at least one gravitational waveform, we very quickly reach a point of exclusion for current
Bayesian algorithms, especially for low mass systems where the length of the waveforms is large and the waveform generation time is on the order of
seconds.  This suddenly demands a timescale of many weeks for a single MCMC.  As each likelihood and Fisher information matrix calculation requires
the evaluation of noise-weighted scalar products, we demonstrate that by using the linearity of integration, and the fact that more than 90\% of the generation time 
is spent at frequencies less that one third of the maximum, we can construct composite integrals that speed up
the MCMCs for comparable mass binaries by a factor of between 3.5 and 5.5, depending on the waveform length.  This method is both source and detector type independent, and can be applied to any waveform that displays  significant frequency evolution,
such as stellar mass binaries with Advanced LIGO/Virgo, as well as supermassive black holes with eLISA.
\\

\end{abstract}

\maketitle

\section{Introduction}
We are entering an exciting era in the field of gravitational wave (GW) astronomy.  ESA has recently chosen the theme of the ``Gravitational Wave Universe" as the theme for the Cosmic Vision L3 mission~\cite{esa}.  Within this framework, a mission
concept called eLISA has been proposed~\cite{elisa}.  This future GW observatory will be composed of three satellites in an equilateral triangle
formation, where each space-craft will be separated by $10^6$ kms.  The four laser links between the satellites will form a single channel Michelson interferometer
which will be sensitive to GWs in the frequency band $10^{-5} \leq f/Hz \leq 1$.  The main sources of interest for this observatory will be the merger of comparable
mass massive black hole binaries, systems where a stellar mass compact object is orbiting a central supermassive black hole (the so-called extreme mass ratio or EMRI
systems), white-dwarf binaries in the Milky Way, and possibly more exotic objects such as cosmic (super)strings and a cosmological background.

Furthermore, upgrades to both Advanced LIGO~\cite{ligo} and Advanced Virgo~\cite{virgo} are in the final stages of development and implementation.  Both GW detectors
should come on-line in the next year or so, heralding a new age of ground-based observations.  It is believed that these detectors will observe many tens of 
stellar mass binaries, composed of neutron stars, white dwarves and black holes.  A newer ultra-low frequency band of GW research is now also evolving via pulsar
timing arrays~\cite{pta}.  As observations improve, it is believed that we should observe at least a background from supermassive black hole binaries, radiating in the
nano-hertz regime before the end of the current decade.
 
While the detectors will have the honour of confirming the first direct detection of GWs from astrophysical sources, the main goal of each project is to use the
recorded events to answer some of the most pertinent questions in astrophysics.  To do this, it is imperative that we are able to estimate the parameters of the 
sources correctly.  In recent years, there has been a shift in GW astronomy from brute-force search and parameter estimation algorithms such as template banks, to more dynamic Bayesian-based algorithms (for an incomplete selection across ground and space-based GW astronomy, see~\cite{B1,B2,B3,B4,B5,B6,B7,B8,B9,B10,B11,B12,B13,B14,B15,B16,B17,B18,B19,B20,B21,B22,B23,B24,B25,B26,B27,B28,B29,B30}).
These algorithms have demonstrated faster convergence to the global solution and reduced run-time.  Furthermore, in the last two decades, the 
Fisher Information Matrix (FIM) was the final word when it came to parameter estimation.  This method provided a quick, and in most cases, a quite accurate
value for the errors in source parameter estimation.  However, in the last decade, especially with the availability of increased computing power, there has been a shift
in philosophy when it comes to parameter estimation.  For the first time in GW astronomy, we are in a position to carry out a full Bayesian inference and map out the 
marginalised posterior densities for the source parameters.  It has been demonstrated that algorithms such as Markov Chain Monte Carlo (MCMC) methods provide a
much more accurate estimation of parameter errors, especially in the case of high parameter correlations and/or low signal to noise ratio (SNR) limit.

As our goal is to extract source signals from the data stream as accurately as possible, GW astronomy is mostly conducted using optimal Weiner or `matched' filtering.
This method is very sensitive to the phase information of the waveform and thus requires a very accurate model of the phase evolution.  In what we shall call the
`standard' method, one is presented with the output of a GW detector, which is assumed to be a linear combination of GW signal and noise.  A theoretical waveform
model or template is then generated, and after transforming to the Fourier domain, is correlated with the data.  This correlation usually involves the calculation of 
scalar quantities such as the SNR or log-likelihood, or a matrix quantity such as the FIM.  In terms of algorithmic development, this is usually a separate process.
An independent template is first generated based on some starting frequency, and a physically motivated ending frequency (e.g. the frequency at the last stable orbit,
the ringdown frequency, or even the upper frequency cutoff of the detector).  This waveform is generated at a constant cadence in order to obey the standard rules of 
Fourier theory for the generation of discrete signals : find the maximum frequency content of the signal, then generate the signal at time-steps given by 
$\Delta t = 1/(N f_{max})$, where $N \geq 2$.   Once this template is generated, it is passed as an array, sometimes with the data array, to another function in order to 
perform an operation (e.g. calculate the SNR etc.).  Seen from this point of view, the template is the primary object of interest.  Once it is generated, all other 
quantities can be calculated.

In this article we present an alternative approach to the problem.  Instead of thinking of the template being the primary object, and all operations as being secondary, 
we suggest inverting the algorithmic structure.  Now it is the operation that is the primary object of interest, and furthermore, it is the operation that dictates how 
a template is generated.  This generic algorithm can be applied to both neutron star and stellar mass black hole binaries in ground-based detectors such as Advanced
LIGO/Virgo, and to SMBHBs in a space-based observatory such as eLISA.  In this work, to demonstrate the advantages of the algorithm, we will use SMBHBs as
our systems of choice.

\subsection{Outline of the paper}
The paper is structured as follows.  In Section~\ref{sec:sec1} we outline the standard methodology for carrying out Bayesian inference for comparable mass binaries and discuss
waveform generation for the calculation of quantities such as the FIM and log likelihood.  In Section~\ref{sec:sec2} we present a method for accelerating 
the calculation of these important quantities via the creation of composite integrals.  As a demonstration, we investigate the acceleration of the log-likelihood and
FIM calculations assuming massive black hole binaries and an eLISA detector.  Finally, in Section~\ref{sec:app} we compare the run-times for a standard method
Hessian MCMC and a composite integral Hessian MCMC.

\section{The Standard Methodology for GW Astronomy}\label{sec:sec1}
\subsection{Bayesian Inference.}
As GW signals are so weak, they are usually buried in the combined instrumental noise of the detector and a galactic foreground.  The main tool currently used for space-based 
GW astronomy is optimal Weiner or matched filtering.  In this case, one takes the output of the detector $o(t) = s(t) + n(t)$, containing
a GW signal $s(t)$ and detector noise $n(t)$, and cross-correlates it with a theoretical waveform model or template $h(t)$.  Matched filtering is very useful for
situations where a weak signal is buried in noise, as it is the optimal linear filter.  We are currently entering an era where the computational power to conduct Bayesian 
inference is becoming more available. Techniques such as Markov Chain Monte Carlo (MCMC)~\cite{MH1,MH2}, Hamiltonian Markov chains~\cite{HMC} and evolutionary
algorithms are becoming more popular, and are starting to replace the more
common Fisher matrix analysis as they allow us to fully map the marginalised posterior density for each parameter.  

As a quick overview of the methodology of the MCMC algorithm : to compare two possible parameter solution sets $x^{\mu}$ and $y^{\mu}$, the main engine  is the Metropolis-Hastings ratio
\begin{equation}
H = \frac{\Pi(y^{\mu})p(s|y^{\mu})q(x^{\mu}|y^{\mu})}{\Pi(x^{\mu})p(s|x^{\mu})q(y^{\mu}|x^{\mu})}.
\end{equation}
Here the quantity $q(x^{\mu}|y^{\mu})$ is a proposal distribution used for jumping from 
$x^{\mu}$ to $y^{\mu}$, $\Pi(x^{\mu})$ are the prior distributions of the parameters and $p(s|x^{\mu})$ is the likelihood defined by 
\beq
p(s|x^{\mu})\equiv{\mathcal L}(x^{\mu}) = C\exp\left[-\frac{1}{2}\left< s-h(x^{\mu}) \left| s-h(x^{\mu})\right.\right>  \right],
\label{eq:likelihood}
\eeq 
where $h(x^{\mu})$ is our waveform model and $C$ is a normalization constant.    The angular brackets define the noise weighted complex inner product of the form
\begin{equation}\label{eqn:scalarprod}
\left<h\left|s\right.\right> =2\int_{0}^{\infty}\frac{df}{S_{n}(f)}\,\left[ \tilde{h}(f)\tilde{s}^{*}(f) +  \tilde{h}^{*}(f)\tilde{s}(f) \right],
\label{eq:scalarprod}
\end{equation}
where a tilde denotes the Fourier transform of the time domain waveform and the quantity $S_n(f)$ is the one-sided noise spectral density of the detector.

For SMBHBs in the high SNR limit, we normally assume that the errors in the estimation of the system parameters are described by a multivariate Gaussian probability distribution 
\beq
p(\Delta x^{\mu}) = \sqrt{\frac{|\Gamma|}{2\pi}}\exp\left[ -\frac{1}{2}\Gamma_{\mu\nu}\Delta x^{\mu}\Delta x^{\nu}\right],
\eeq 
where we define the FIM
\beq
\Gamma_{\mu\nu} = \left<\frac{\partial h}{\partial x^{\mu}}\left|\frac{\partial h}{\partial x^{\nu}}\right. \right> = -E\left[ \frac{\partial^2 \ln {\mathcal L}}{\partial x^{\mu} \partial x^{\nu}}\right],
\label{eq:FIM}
\eeq
as the negative Hessian on the log-likelihood, and $|\Gamma| = det(\Gamma_{\mu\nu})$.   If the value of the Metropolis-Hastings ratio is superior to a uniform random draw $U[0,1]$, the chain
then moves to the next point in parameter space with a probability $\alpha = min(1,H)$, otherwise the chain stays at $x^{\mu}$ and proposes a new point.  If we were
to use what is commonly known as a Hessian MCMC to make jumps in the
parameter space, the most efficient proposal distribution to use is a multivariate Gaussian based on the FIM.  The Hessian MCMC is a popular choice of algorithm as it has been shown to have the highest acceptance rate for this class of random walk algorithms.  However, while a Hessian MCMC boasts the highest acceptance 
rate for random walk MCMCs, it can still be quite low, requiring a long chain to converge to the target density.  In general for SMBHBs, one would like to run MCMCs
that have at least $10^6$ iterations.  This means that even if we only calculate the FIM every 100 iterations, we still need $10^4$ FIM and $10^6$ log likelihood
evaluations per chain.  In terms of waveform numbers, this translates to $1.7\times10^5$ waveforms for the FIM calculation (17 waveforms are required per FIM
due to the fact that we need to calculate numerical derivatives of the response function) and $10^6$ waveforms for the log likelihoods.  Once the time for a waveform generation approaches the order of one second or more, this means that a $10^6$ iteration MCMC will take approximately 16-17 days to compute (including additional
overheads involved in the algorithm).  This implies that at present, we are unable to carry out a Bayesian inference for low mass systems, whose waveform sizes
are so large that the generation time is anything up to 10 seconds per waveform (a timescale that would result in the MCMC taking approximately four months to run).

More recently it has been demonstrated that a non-random walk variant of the MCMC, called a Hamiltonian Markov Chain (HMC)~\cite{HMC}, has a superior convergence rate
to a Hessian MCMC resulting in a reduced run-time for Bayesian inference~\cite{B24}.  However, this work also demonstrated that while the main bottleneck for a Hessian MCMC
is the calculation of the likelihood and FIM, the bottleneck for the HMC is the calculation of the likelihood and gradients of the log-likelihood.  One of the main results
of Ref~\cite{B24} was that while the HMC converges approximately the dimensionality of the problem faster than a Hessian MCMC, it requires approximately 1.5 times
as many waveform generations.

\subsection{Waveform Generation}
Once our method for Bayesian inference is chosen, a choice of waveform model rests.  At present for SMBHBs, there are no complete inspiral, merger and ringdown
waveforms  encompassing all of the important physical effects (i.e. spin, eccentricity, higher harmonics etc).  It has been shown that inspiral waveform models incorporating higher harmonic 
corrections (HHC) have the capability of breaking correlations between parameters and improving parameter estimation~\cite{hhc1,hhc2,hhc3,hhc4,hhc5}.  Because of this we will choose
a 2PN HHC waveform as our model for the GW polarisations~\cite{biww}
\begin{equation}
 h_{+,\times}  =  \frac{2 G m \eta}{c^2 D_L} x\left[ H^{(0)}_{+,\times} + x^{1/2} H^{(1/2)}_{+,\times}  + x H^{(1)}_{+,\times} + x^{3/2} H^{(3/2)}_{+,\times} + x^2 H^{(2)}_{+,\times} \right].
 \label{eqn:hhcwave}
\end{equation}
The polarisations $h_{+,\times}(t)$ are functions of the total mass of the binary $m=m_1 + m_2$, the reduced mass ratio $\eta = m_1 m_2 / m^2$, the luminosity distance to the source $D_L$, and the post-Newtonian velocity parameter $x(t)$ which is a function of the binary orbital frequency $\omega(t)$.   The functions $H^{(n/2)}_{+,\times}(t)$ are the post-Newtonian higher harmonic correction terms which are functions of the individual masses, the inclination of the binary along the line of sight $\iota$, the time to coalescence of the
system $t_c$, the sky position of the source $(\theta,\phi)$ and integer multiples of the orbital phase.  The total response from a GW observatory is then written as
\begin{equation}
h(t) = h_{+}(\xi(t))F^{+}(t)+h_{\times}(\xi(t))F^{\times}(t). 
\label{eq:hoft}
\end{equation}
where $t$ is the time in the solar system barycenter, $\xi(t)$ is a Doppler-induced phase shifted time parameter due to the motion of the detector, and the functions $F^{+,\times}(t)$ are the beam pattern functions of the detector given in the low frequency approximation by~\cite{RPC,cc}
\beq
F^{+}(t;\psi, \theta, \phi, \lambda) = \frac{1}{2}\left[\cos(2\psi)D^{+}(t;\theta, \phi, \lambda) - \sin(2\psi)D^{\times}(t;\theta, \phi, \lambda)\right],
\eeq
\beq
F^{\times}(t;\psi, \theta, \phi, \lambda) = \frac{1}{2}\left[\sin(2\psi)D^{+}(t;\theta, \phi, \lambda) + \cos(2\psi)D^{\times}(t;\theta, \phi, \lambda)\right],
\eeq
where $\psi$ is the polarization angle of the wave and $\lambda = 0$ or $\pi/4$ depending on whether we have a single or two channel detector.  The expressions for the detector pattern
 functions $D^{+,\times}(t)$ will not be repeated here, but can be found in~\cite{RPC}

As the GW analysis is carried out in the Fourier
domain, it is necessary for us to employ a numerical Fast Fourier Transform (FFT) to convert the time domain waveform. Therefore, when generating the above time-domain waveforms we need to take care of the chosen sampling period $\Delta t$.   In standard Fourier
theory, a time domain signal should have a sampling period which is at least twice the Nyquist frequency, i.e. the highest frequency component of the wave, to prevent
 aliasing.  Aliasing occurs in poorly sampled data when power outside of the frequency range of interest is moved into this range, giving spurious signal power
in the Fourier domain.  If we take the upper
frequency cutoff of either the detector, or the waveform, as the Nquist frequency, the sampling rate in the time domain is given by $\Delta t = 1/f_s$, where $f_s = (2f_{Nyq}) = 2f_{max}$ is the sampling frequency.  In reality, we use a sampling frequency of $f_s\geq4f_{max}$ just to be certain that we avoid aliasing.  So, defining our sampling frequency as $f_s = 4f_{max}$ for a signal of duration
$T$, the number of points $n$ needed, as a power of 2,  to represent the time domain waveform is given by
\beq
n = 2^{\lfloor(log_x(f_s T) / log_x(2))+1\rfloor}
\eeq
where $x$ defines a logarithmic base of choice and the brackets $\lfloor\rfloor$ denote the floor symbol.  Now, as the GW frequency is inversely proportional to 
the total redshifted mass, i.e. $f_{GW} = x^{3/2} / (\pi m(z))$, the smaller the total mass of the system, the higher the maximum frequency of the waveform, and
hence the larger the value of $n$.

Given the previously defined waveform model, for systems that have a time to coalescence $t_c$ which is less than the observation time of the data $T_{obs}$, we terminate
the waveform at a minimum separation between the SMBHs of $R=7M$ (where we assume $G=c=1$).  For our analysis we normally choose a minimum separation of $R=7M$ rather than the 
separation at the last stable orbit $R=6M$ to terminate the waveforms.  This is mostly due to the fact that for some combinations of $(m,\eta)$, the gradient of the
orbital frequency changes sign before reaching the last stable orbit.  This sudden change of sign implies an unphysical outspiral of the system.  We have verified
that the evolution of $\omega(t)$ remains monotonic down to a separation of $R=7M$. With this is mind, we thus define the maximum frequency component of the 
dominant harmonic as
$f^{22}_{max} = f^{22}(R=7M)$.  For the 2PN HHC waveform, the highest harmonic radiates at six times the orbital frequency, or three times the 
GW frequency, meaning that $f^{HHC}_{max} = 3 f^{22}_{max}$.    For systems where $T_{obs} > t_c$, we calculate the minimum separation $R_{min}=R(t=T_{obs})$
and repeat the 
process by first finding $f^{22}_{max} = f^{22}(R=R_{min})$.

As an aside, we should mention that for systems where we evolve the waveforms right to the point of minimum separation, the choice of upper frequency cutoff in 
the inner product integral is different for the calculation of the likelihood and the FIM.  For the likelihood, we are comparing a template with a possible signal.  If the 
search phase of our algorithm has worked properly, we should be carrying out the inference study at the global peak in the multi-dimensional likelihood surface.  To 
ensure that
we are exploring all sides of the marginalised posterior, we need to make sure that we are not restricting the inference to one side of the peak by only integrating
the likelihood to the maximum frequency of the best fit template.  If, for example, we had a perfect template match in the high SNR regime, i.e. $h(t)=s(t)$, this integration frequency would correspond to the central point of the likelihood peak.  If we take this value as an absolute cutoff in the integration, it would mean that
only templates with a total mass greater than the maximum fit template would be generated to $R=7M$, while lower total mass systems would be truncated
earlier and earlier as the minimum separations between SMBHs would be at best $R > 7M$.  This asymmetry in the waveform generation would introduce a bias
in the Bayesian inference.

Therefore, we normally integrate the likelihood function to a frequency of $f_{int} = 2f^{HHC}_{max}$.  This guarantees that lower mass systems are also fully taken
into account and the likelihood peak is fully explored.  Furthermore, when it comes to defining the sampling frequency, we then define $f_s=2f_{int}=4f^{HHC}_{max}$,
ensuring that the Nyquist theorem is respected.  On the other hand, the FIM calculation is a pure template calculation, i.e. $\Gamma^{\mu\nu} = <\partial_{\mu}h|\partial_{\nu}h>$ where $\partial_{\mu}=\partial / \partial x^{\mu}$.  In this case, we simply integrate the inner product over waveform derivatives to the maximum cut-off frequency of the waveform, but again respect the required sampling rate.

With all of this in mind, and taking the lower limit of integration in Eqn~(\ref{eqn:scalarprod}) to be the low frequency cutoff an an eLISA-type detector, i.e. $10^{-5}$ Hz, we
can redefine both the log-likelihood and FIM integrals as
\begin{equation}
\ln {\mathcal L} =\left< s-h \left| s-h\right.\right> = 4\int_{10^{-5}}^{2f^{HHC}_{max}}\frac{df}{S_{n}(f)}\,\left| \tilde{s}(f) - \tilde{h}(f)  \right|,
\label{eq:scalarprod2}
\end{equation}
and 
\begin{equation}
\Gamma^{\mu\nu} = 4\int_{10^{-5}}^{f^{HHC}_{max}}\frac{df}{S_{n}(f)}\,\frac{\partial \tilde{h}(f)}{\partial x^{\mu}}\frac{\partial \tilde{h}^{*}(f)}{\partial x^{\nu}}
\end{equation}
where the pre-factor increases because of the contribution from the complex conjugate term. 

\section{Accelerating GW Algorithms}\label{sec:sec2}
There have recently been a small number of papers which have investigated accelerating the likelihood calculation, with a view to reducing the run-time of Bayesian style
algorithms.  These algorithms suggest using alternative techniques such as a heterodyning~\cite{acc1}, likelihood transformations~\cite{acc2}, reduced
order methods~\cite{acc3} or surrogate waveform models~\cite{acc4,acc5}.   We add to this same sample by suggesting a new approach to the problem.

\subsection{Methodology}
\begin{figure}[t]
\begin{center}
\epsfig{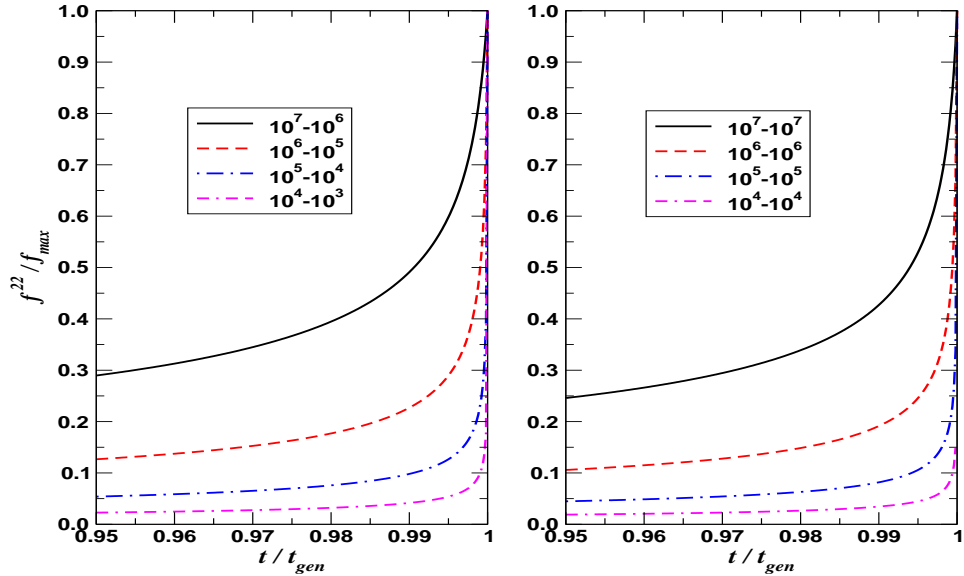}
\end{center}
\caption{Fractional evolution of the dominant GW frequency as a function of fractional waveform generation time for systems with mass ratio $q=10$ (left)
and mass ratio $q =1$ (right) for sources at $z=1$.  We can see that in all cases, when generating gravitational waveforms we spend more than 95\% of the time at frequencies less than
$0.3 f_{max}$.}
\label{fig:evolution}
\end{figure}

We have seen in the previous section that to avoid aliasing, one needs to sample the time domain waveform at multiples of the Nyquist or highest frequency
of the waveform.  Especially for low mass or long lived systems, this can lead to a waveform generation that is computationally intensive and can result in 
a computational bottleneck if we are hoping to carry out a Bayesian inference. To explain the main idea behind the acceleration technique, we draw the readers attention to Fig~\ref{fig:evolution}.  Here we plot the dominant 22-mode frequency evolution as a fraction
of the maximum frequency of the waveform, against the waveform generation time as a fraction of the total generation time, i.e. $f/f_{max} \,vs.\, t/t_{max}$.  On the left
hand side we plot a range of source frame masses with mass ratio $q=10$, while on the right hand side we plot a set of systems with mass ratio $q=1$.  In each case the source is placed at a redshift of $z=1$.

We can see that for the highest mass systems , i.e. $(10^7,10^6)$ and $(10^7,10^7)\,M_{\odot}$, 95\% of the waveform generation time is spent at frequencies lower
than $0.3f^{22}_{max}$.  As we go to lower masses, this timescale increases and the waveforms can actually spend up to 99\% of the waveform generation times at frequencies less than
$0.2f^{22}_{max}$. This implies that the majority of waveforms, in the standard method of analysis and especially if HHCs are included, are grossly oversampled.  With this in mind, and using the linearity of the inner product, our idea is to break up all inner product integrals such as Eqn~(\ref{eqn:scalarprod}) into two parts, i.e.
\begin{equation}
\left<s\left|h\right.\right> = \left<s\left|h\right.\right>_L + \left<s\left|h\right.\right>_H
\end{equation}
where the low frequency integral is defined by
\begin{equation}
\left<s\left|h\right.\right>_L =2\int_{10^{-5}}^{f(t_{x\%})}\frac{df}{S_{n}(f)}\,\left[ \tilde{s}(f)\tilde{h}^{*}_L(f) +  \tilde{s}^{*}(f)\tilde{h}_L(f) \right],
\label{eq:scalarprod2}
\end{equation}
and the high frequency integral is given by
\begin{equation}
\left<s\left|h\right.\right>_H =2\int_{f(t_{x\%})}^{f_{max}}\frac{df}{S_{n}(f)}\,\left[ \tilde{s}(f)\tilde{h}^{*}_H(f) +  \tilde{s}^{*}(f)\tilde{h}_H(f) \right].
\label{eq:scalarprod2}
\end{equation}
The high/low integration limit $f(t_{x\%})$ denotes the dominant harmonic frequency at a time when $x\%$ of the waveform has been generated.  We will
discuss how $t_{x\%}$ is chosen in the upcoming sections.  The advantage of breaking the integrals into low and high frequency integrals is that the waveforms
needed for each integral can now be generated at different sampling rates.  The first integral requires the generation of a waveform $h_L(t)$ at a cadence of $\Delta t \geq 4f(t_{x\%})$, where this waveform accounts for $x\%$ of the total waveform.  The waveform for the second integral $h_H(t)$ is generated at the standard rate of $\Delta t\geq4 f_{max}^{HHC}$.  However, while the sampling rate is higher, this waveform only accounts for $(100-x)\%$ of the total waveform.  While this method
now requires two independent FFTs for both $h_L(t)$ and $h_H(t)$, the generation time saved due to the reduced sampling period for $h_L(t)$ is more than sufficient
to compensate for a second FFT and still provide an overall speedup.

\subsection{Reducing the HHC content of the low frequency waveform}
To obtain a further reduction in the time taken to calculate the log-likelihood, we again draw attention to the fact that in the standard method, if we include the 
HHCs, the maximum frequency content of the waveform is $f_{max}^{HHC} = 3f_{max}^{22}$.  We can therefore think of separating the power spectrum of a waveform into three distinct frequency bands.  In
the medium band, which encompasses the majority of the region of interest, we have the power due to the dominant $f^{22}$ harmonic.  From a few times $10^{-5}$Hz and 
up to a frequency of $f^{22}(7M)$ (hereafter $f_{7M}$ for brevity), this harmonic swamps all other sub-dominant harmonics and is the main contributor to the total 
power.  In fact if one calculates the log-likelihood or FIM error estimation for a source by only integrating to a maximum frequency of $f_{int}=f_{7M}$, the results
are almost identical regardless of whether the HHCs are included or not.  In fact the small discrepancy in results is due to the structure at low frequencies coming
from the $f=f_{orb}=1/2 f^{22}$ harmonic.  We can thus think of the small frequency range where the structure from this harmonic is visible as being the low
frequency band in our catagorisation.

In the high frequency band, it is only at frequencies greater than $f_{7M}$ that the higher sub-dominant harmonics become truly visible.  In this band between
$f_{7M} \leq f/Hz \leq f^{HHC}_{max}$, we gain the further information that breaks parameter correlations and improves accuracy.  We should point out that 
harmonics contained in the $H^{(2)}_{+,\times}(t)$ function radiate at frequencies of $f=6f_{orb}=3f^{22}$, meaning that for HHCs to be visible in this high band, one
has to start generating full waveforms at frequencies no greater than $f=1/3f^{22}$ at the most.  With this information in mind, it became clear that a maximum low
frequency of $f_L^{max} = 1/3f^{22}$ could be taken as a possible termination criterion for the low frequency waveform.  This would now imply that the sampling
frequency for this waveform could be set at $f_s =4f_L^{max} = 4/3f^{22}$, as opposed to $f_s = 4f^{HHC}_{max} = 12f^{22}$ in the standard case.  Furthermore, 
this choice would mean that it would no longer be necessary to generate the full compliment of HHCs for the low frequency waveform as we are never integrating
higher than this termination frequency.

One choice would be to generate $h_L(t)$ by just using a restricted PN waveform, i.e. only retaining the $H^{(0)}_{+,\times}(t)$ term in Eqn~(\ref{eqn:hhcwave}). 
However, to account for the additional  low frequency structure appearing due to the $f=1/2 f^{22}$ harmonic, we decided to use a 0.5PN HHC waveform model, i.e.
\begin{equation}
 h_{+,\times}  =  \frac{2 G M \eta}{c^2 D_L} x\left[ H^{(0)}_{+,\times} + x^{1/2} H^{(1/2)}_{+,\times} \right], 
\end{equation}
to generate $h_L(t)$.  Here,  the factors $H_{+,\times}^{(1/2)}$ introduce harmonics at frequencies of $f=1/2f^{22}$ and $f=3/2f^{22}$.  This single correction gives 
enough structure to the waveform at these frequencies that it is an excellent match to the full 2PN HHC waveform in this band.

\subsection{Choosing a termination point}
As the most important quantity arising in the Metropolis-Hastings ratio is the likelihood function, to test where best to terminate the first, and begin the second integral, 
we decided to investigate the error in the calculation of the log-likelihood, by requiring that $\ln{\mathcal L}_L + \ln{\mathcal L}_H \approx \ln{\mathcal L}$ for 
different values of $f(t_{x\%})$.  On investigating a single source, we found that it is vitally important that both low/high frequency waveforms are properly windowed at both the beginning and the end of
 each waveform generation.  Failure to do so results in an excess power phenomenon called ``ringing" in the Fourier domain, which results in spurious values for the individual log-likelihood calculations.  To properly deal with this issue, we begin and end each waveform using a half-Hann window to reduce excess power in the Fourier domain to an acceptable level.

In the previous section, we suggested a possible termination frequency of $f_{int} = 1/3f_{7M}$ (hereafter $f_{1/3}$) for the low frequency waveforms (and clearly
imposing an starting frequency for the high frequency waveforms).  This initially seems like a valid choice as only HHCs generated above this frequency are visible
at frequencies above $f_{7M}$.  However, if we refer to Fig~\ref{fig:evolution}, we can see that many of the low mass systems do not cross this threshold until the 
final 5\% of the waveform generation time.  Similarly for systems with a total redshifted mass of $m(z)\geq2\times10^7\,M_{\odot}$ and $q\geq10$, this frequency is 
reached earlier in the waveform generation.  While this shift is not important for the high mass systems, it is a real problem for low mass waveforms.  In general, 
the low frequency integral is the minor contributor to the total integral.  Therefore, if the high
frequency waveform is too short, its power is effectively killed by the windowing at the beginning and end of the waveform.  The result of this is a gross error in the 
high frequency integral, meaning that we do not come close to the true log-likelihood value upon addition of the composite integrals.  This further implies that there is a minimal critical length to the high frequency waveform if we want to properly approximate the full log-likelihood calculation.

To investigate this point further, we decided to calculate the error in the log-likelihood by using either $f_c=min\{f_{1/3}, f(t_{x\%})\}$, as the point of 
intersection in frequency space, where we used the range $0.9\leq t_{x\%}/t_{max}\leq 0.99$.  Our first calculations, even at $t_{x\%} = 0.9 t_{max}$, demonstrated
errors of greater than 10\% between the two methods of calculation.  Upon further investigation, we found that the two waveform power spectra from the composite 
calculations did not perfectly overlap the power spectrum of a full waveform.  The reason for this was the smooth rise and fall-off in the spectra, due to the 
window function, introduced a tapering effect that left a gap in the power spectrum.  This meant that the composite inner products were underestimating the true value.
To circumvent this problem, we decided to generate the low frequency waveform to a termination frequency of $f = f_c+\delta f$, and to begin the high frequency
waveform generation at $f = f_c - \delta f$, where $\delta f$ is chosen to be sufficiently large such that the window function has a value of unity at $f=f_c$.  This
negates the tapering effect of the window function in the integrals and returned an error of $< 10^{-2}$\% in the log-likelihood calculation for the particular source
in question.

While this single source test demonstrated that the algorithm was working, to obtain a more global picture, we ran a Monte Carlo simulation of $3\times10^4$
sources from an astrophysical distribution~\cite{sesana}.  The time to coalescence was chosen to lie between $0.2\leq t_c/yrs\leq 0.99$, inclination and co-latitude
were each chosen from flat distributions in $\cos\iota$ and $\cos\theta$, while all other parameters were chosen randomly from flat distributions in their natural 
values.  Furthermore, for an eLISA-type detector, if we run a null-signal test, i.e. $o(t)=n(t)$, it can be shown that a SMBHB waveform can find points in Fourier
space where it has a non-zero correlation with random noise fluctuations.  This null-signal test can be used to set a detection threshold for a matched-filtering
algorithm.  We have recently demonstrated that for a single-channel eLISA observatory, a MCMC conducted using non-spinning quasi-circular HHC waveforms
sets a detection threshold of $\ln {\mathcal L} \approx 50$ (or $\rho\approx10$).  As a measure of success, we demand that the composite log-likelihood calculates
incur an error of approximately 1\% when compared to the standard version.

\begin{figure}[t]
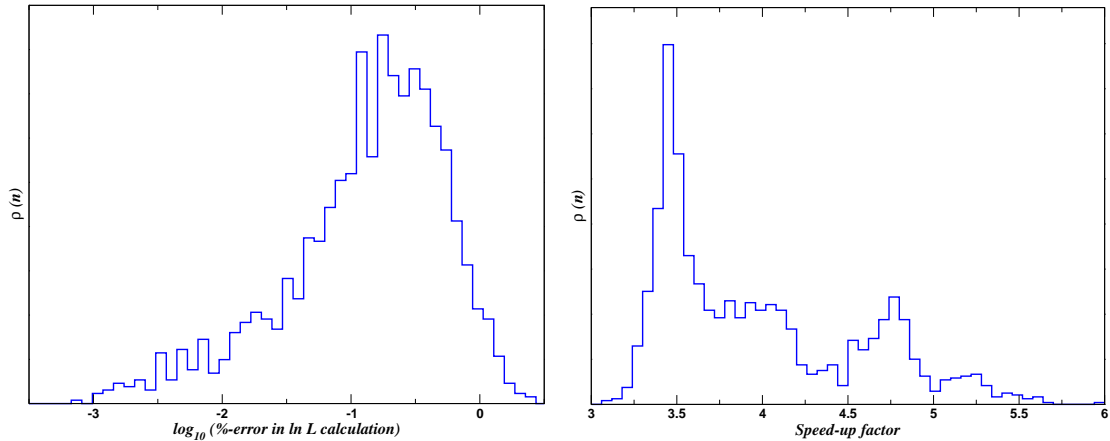

\minipage{0.4\textwidth}
  \includegraphics[width=\linewidth]{lnL_Error_Histogram.eps}
\endminipage\hspace{0.2cm}
\minipage{0.4\textwidth}
  \includegraphics[width=\linewidth]{lnL_Speedup_Histogram.eps}
\endminipage\hfill
  \caption{On the left we plot the percentage error between the full and combined log likelihood calculations.  We note that the majority of sources return an
  error of less than 1\%.  On the right we plot the gain in calculation speed between the full and combined log likelihood calculations.}
  \label{fig:histograms}
\end{figure}

We ran Monte Carlo simulations for the different values of $t_{x\%}$ at 0.01 interals and primarily investigated both the percentage error and the speed-up factor
in the calculations.  In general, for $t_{x\%} \leq 0.94t_{max}$, the errors were much smaller than the 1\% required, but the speed-up factors were minimal (1.4-1.7)
over the full calculation.  For $t_{x\%} \geq 0.96t_{max}$, the errors now began to grow to as much as 40\%, while the speed-ups were impressive at factors of
8-12. In Fig~\ref{fig:histograms} we present normalised histograms representing the percentage error in the log-likelihood calculations (left panel) and the 
speed-up factor for the composite integrals over the full calculation (right panel), assuming a threshold of $\ln {\mathcal L} \geq 50$, for the case of $f_{int}=min\{f_{1/3}, f(t_{x\%}=0.95t_{max})\}$.  We can see that in the vast majority of cases, the percentage error in the log-likelihood calculation is less than 1\%.  In fact, the
systems that returned errors of greater than 1\% were those at the detection limit.  From the right hand panel, we observe that the majority of systems achieved
a speed-up factor of $\sim3.5$, with many systems enjoying speed-up factors of anything up to 6.  The peak of events with a speed-up of $\sim3.5$ is populated
by lower mass sources, i.e. redshifted total masses of $m(z) \leq 10^6\,M_{\odot}$, which require array sizes of $\geq2^{19}$.  The systems that achieve the greatest 
acceleration are those with redshifted total masses of $m(z) \geq 3\times10^6\,M_{\odot}$ and array lengths of $2^{16}\leq 2^n \leq 2^{18}$.  Unfortunately, an 
acceleration of 3.5 may already be the maximum achievement for low mass sources.   The success of separating the integrals into low/high frequency bands depends on a significant 
frequency evolution of the source, (i.e. on the level of one order of magnitude over the entire duration of the waveform).  For low mass systems, from when they are first visible in 
the detector, they are already at quite high frequencies (i.e. a few mHz).  They essentially stay at this frequency, with a painfully slow evolution, right until the final final stage of evolution when the frequency then changes dramatically.  Because of this essentially monochromatic behaviour,  the acceleration is minimal.  We will
investigate other means  of acceleration for these systems in the future.

\subsection{Accelerated Fisher Matrices}
We can apply the same techniques developed for the log-likelihood approximation to the calculation of the FIM.   By using the additive properties of the Fisher matrix, we can again divide the calculation into  
low/high frequency integrals, i.e.
\begin{equation}
\left< \partial_{\mu} h | \partial_{\nu} h \right> = \left< \partial_{\mu} h | \partial_{\nu} h \right>_L+\left< \partial_{\mu} h | \partial_{\nu} h \right>_H.
\end{equation}
As with the log-likelihood calculation, we again use the frequency $f_{int}=min\{f_{1/3}, f(t_{x\%}=0.95t_{max})\}$ as the upper/lower limit of integration.  With the 
FIM we can be a lot more relaxed in the accuracy of the approximation.  In general it is the eigenvalues and eigenvectors of the FIM that are used for both the 
magnitude and direction of the jump in parameter space.  Therefore, as long as these are within a factor of two of the full FIM calculation, there is no apparent 
loss of acceptance rate (also due to the fact that the magnitudes of the jump proposals are scaled with a random Gaussian pre-factor and the inverse dimensionality of the search space).  Ultimately,
our choice of integration frequency results in errors of $\sim1\%$ in the eigenvalues, rising to a around 10\% for the more massive systems, making the choice more than acceptable.

\begin{figure}[t]
\begin{center}
\epsfig{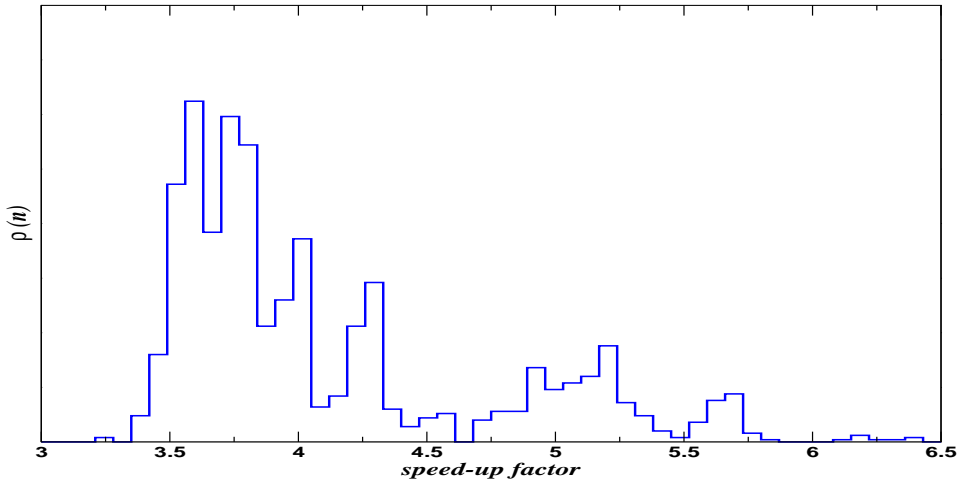}
\end{center}
\caption{A histogram demonstrating the speed-up factor in the FIM calculations by using a composite low/high frequency methodology  The slowest acceleration
factors are for the systems which require the longest waveforms.}
\label{fig:fimacc}
\end{figure}

To test the acceleration of the FIM calculation, we again ran a Monte Carlo simulation using the same guidelines as in the previous section.  In Fig~\ref{fig:fimacc} we
plot a histogram of the acceleration factors achieved by using the composite integral method.  The image present a similar picture to what we saw with the log-likelihood calculation.  We obtain a peak acceleration of between 3.5-4 for a large number of sources, with a second smaller peak at an acceleration factor of $\sim5$,
and a smaller number of sources enjoying an acceleration factor of anything up to 6.5.  Once again, the smallest speed-up factor corresponds to the lower mass
systems in our distribution, which require the longest waveform generations, and thus, the largest array sizes.

\section{Application to a Hessian MCMC}\label{sec:app}
\begin{figure}[t]
\begin{center}
\epsfig{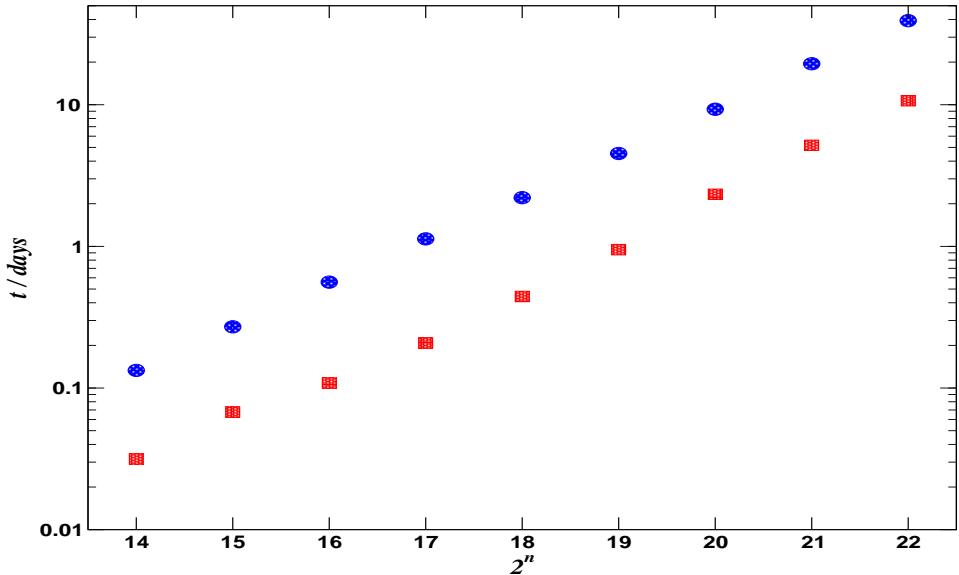}
\end{center}
\caption{The average run-time in days for a standard method MCMC (blue circles) and the composite integral method MCMC (red squares) for a $10^6$ iteration
algorithm.  At low array sizes, the composite method achieves a speed-up factor of 4.  The maximum speed-up factor is 5.44 for array sizes of $2^{17}$, dropping
off to a factor of 3.7 for array sizes of $2^{22}$.}
\label{fig:mcmcalc}
\end{figure}

While the two previous sections have demonstrated an acceleration in both the log-likelihood and FIM calculations, what we are really interested in is how this 
translates to a saving in run-time for a full algorithm.  To test this, we ran a number of Hessian MCMCs for different mass systems, requiring different waveform lengths
from $2^{14}$ to $2^{22}$.  In each case, we assumed an observation period of one year, with the time to coalescence again chosen randomly between $0.3\leq t_c/yrs\leq 0.99$.
Each algorithm was run for $10^6$ iterations, with a FIM update every 100 iterations.  The algorithms were run on a quad-core MacBook Pro equipped with a 2.6 GHz Intel i7 CPU and 8GB of memory.  In Fig~\ref{fig:mcmcalc} we plot the average time taken in days for both a 
standard MCMC (blue-circles) and the composite-integral MCMC (red-squares), as a function of array size as a power of 2.

As the MCMC is a linear algorithm, doubling the number of array points approximately doubles the run-time of the algorithm.  Similarly, doubling the number of 
iterations also approximately doubles the run-time of the algorithm.  This explains the almost linear behaviour of the run-time as a function of array size in the
log-linear plot.  In the standard method, we can see that algorithms with array sizes of $\leq2^{16}$ are relatively quick, always taking less than a day to run for a 
$10^6$ iteration MCMC.  However, for the majority of the sources in the eLISA band, we usually require array lengths of $\geq2^{17}$.  In this case, the algorithm
takes approximately one day to run.  As we increase the array size, the run-time effectively doubles for each power of 2, reaching an average run-time of 40 days
for a waveform size of $2^{22}$.  This scale of run-time is clearly prohibitive as a technique for estimating the parameters of a system, especially if the results are
needed quickly to form the priors for the next run. 

On the other hand, by using the composite integral method, we can see that we achieve run-time accelerations of between 4-5 at array sizes of $\leq2^{16}$.  While 
interesting, our main concern is the acceleration of the algorithm when it takes greater than a day to run.  To this end,  we observe that for the $2^{17}$ case, we now 
reduce the run-time from 1.13 days to $\sim5$ hours, a speed-up of 5.44.  For the $2^{20}$ case, the run-time is reduced from $\sim10$ days to 2.33 days (speed-up
of $\sim4$), and for the $2^{22}$ case, we reduce the run-time from a crippling 40 days to 10.6, giving and acceleration factor of $\sim3.7$.

We should highlight once more that this algorithm is independent of source and detector type.  The use of composite integrals produces an acceleration factors, for
Bayesian inference MCMC algorithms, of 4 for small array lengths ($\leq2^{16}$), peaking at 5.44 for array lengths of $2^{17}$, and drops of to a factor of 3.7 for 
array sizes of $2^{22}$.  This acceleration, while demonstrated here for SMBHBs with the eLISA observatory, is also applicable to neutron star and stellar mass
black hole binaries using ground-based detectors such as Advanced LIGO/Virgo.

\section{Conclusion}
We have introduced a new approach to how algorithms are designed for conducting GW astronomy using comparable mass binary systems.  The standard method
of design is to first generate a template, then use that template to evaluate an operation such as a log-likelihood or SNR calculation.  Done this way, this 
requires the template to be generated at a constant cadence.  Especially when higher harmonic corrections are introduced, and also due to the fact that the majority
of the waveform generation time is spent at frequencies lower that a third of the maximum waveform frequency, this means that standard templates are grossly
oversampled.

By making the operation the primary concern, and template generation secondary, we demonstrated that it is possible to achieve a speed-up in algorithm run-time
by re-designing the algorithmic structure.  Now it is the operation that dictates how the waveform should be generated.  By using the linearity of integration, we 
demonstrated that it is possible to split all log-likelihood and FIM integrals into low and high frequency components.  This allows us to use a short low-cadence waveform for the low frequency integral, and a much shorter high-cadence waveform for the high frequency integral.  As well as requiring smaller waveform sizes, we
also demonstrated that we could achieve a further acceleration by reducing the PN order of the higher harmonic corrections in the low frequency integral without 
incurring a large error.  When applied to a $10^6$ iteration Hessian MCMC, where the FIM is updated every 100 iterations, it allowed us to reduce run-times for waveform sizes of : $2^{14}$ by
a factor of 4 (3.2 hours to 45 minutes),  $2^{17}$ by a peak factor of 5.44 (1.13 days to 5 hours) and  $2^{22}$ by a factor of 3.7 (40 days to 10.6 days).

While we demonstrated the advantages of this method for the investigation of SMBHBs with the future eLISA observatory, this method is source and detector 
independent, making it just as applicable to neutron star and stellar mass black hole binaries using Advanced LIGO/Virgo.

\end{document}